\documentclass[%
 aip,
 jmp,%
reprint,%
 byrevtex,
]{revtex4-1}

\usepackage[latin1]{inputenc}
\usepackage{times}
\usepackage{mathptmx}
\usepackage{bm}
\usepackage{microtype}
\usepackage{mathtools}
\usepackage{comment}
\usepackage{color}
\usepackage{booktabs}
\usepackage{graphicx}
\usepackage{dcolumn}

\RequirePackage{xspace}

\newcommand*{\cf}{\emph{cf.}\xspace}

\newcommand*{\eg}{\emph{e.g.}\xspace}

\newcommand*{\ie}{\emph{i.e.}\xspace}

\newcommand*{\eqn}[1]{equation~(\ref{#1})}
\newcommand*{\eqns}[1]{equations~(\ref{#1})}

\newcommand*{\eps}{\varepsilon}
\newcommand*{\half}{\frac{1}{2}}
\newcommand*{\im}{{\mathrm{i}}}

\renewcommand*{\d}[1]{\mathrm{d}#1}
\newcommand*{\ddz}[1]{\frac{\d{#1}}{\d{z}}}
\newcommand*{\ddzz}[1]{\frac{\d{^2#1}}{\d{z^2}}}
\newcommand*{\ddZ}[1]{\frac{\d{#1}}{\d{\zeta}}}
\newcommand*{\ddZZ}[1]{\frac{\d{^2#1}}{\d{\zeta^2}}}

\begin{document}

\title[New Phase-Integral Method Platform Function]%
{New Phase-Integral Method Platform Function}

\author{S.\,Yngve}
 \email[Electronic mail: ]{staffan.yngve@fysast.uu.se.}
 \affiliation{%
  Department of Physics and Astronomy,
  Theoretical Physics,
  Uppsala University,
  P.\,O.~Box~516,
  SE-751\,20~Uppsala,
  Sweden,
 }

\author{B.\,Thidé}
 \email[Electronic mail: ]{bt@irfu.se.}
 \homepage[Home page: ]{www.physics.irfu.se/~bt.}
 \affiliation{%
  Swedish Institute of Physics,
  Physics in Space,
  P.\,O.~Box~537,
  SE-751\,21~Uppsala,
  Sweden
 }
 \altaffiliation[Also at ]{%
  LOIS Space Centre,
  Växjö University,
  SE-351\,95~Växjö,
  Sweden.%
  }

\date{12 December 2009}
\revised{\today}

\begin{abstract}

The phase-integral method (PIM) is an asymptotic method of the
geometrical optics or semi-classical type for solving approximately, but
in many cases very accurately, a wide class of differential equations in
physics. Unlike the related (J)WKB method, the higher-order corrections
in the PIM can be generated from a generic, unspecified base function,
providing added symmetry and flexibility. However, with the conventional
approach of using the next-to-lowest (third) order correction to the
integrand in the phase integral as a platform for calculating higher
(fifth, seventh, ninth,\ldots) order corrections, the higher-order
calculations very often become quite complicated.

We therefore introduce a new platform function, which considerably
simplifies the calculation of the third-order contribution for a wide
range of problems. We also present directly integrable conditions for the
phase integral, which so far seem to have gone unnoticed.

For a large number of observables, our analysis makes a clearer
distinction between physical and, in a sense, unphysical contributions.

\end{abstract}

\maketitle

\section{Introduction}

In their pioneering 1936 article on the calculation of Coulomb wave
functions in non-relativistic quantum mechanics, Yost, Wheeler and Breit
\cite{Yost&al:PR:1936} commented on the fact that the WKB method in the
first-order approximation sometimes has an unsatisfactory accuracy.
Moreover, they found that the replacement, originally suggested by
Kramers,\cite{Kramers:ZP:1926} of $l(l+1)$ by $(l+1/2)^2$ where
$l$ is the orbital angular momentum quantum number, does not always
improve the results obtained in the first-order WKB approximation.
This situation can mostly be remedied by going to the next-order
approximation. The phase-integral method of Fröman and Fröman
\cite{Froman&Froman:Book:1965} provides a straightforward structure
for calculating the contributions of order five and higher. For this,
Ref~\onlinecite{Froman&Froman:Book:2002} introduces a platform function
$\eps_0(z)$ which, apart from a factor $1/2$, is the third-order
contribution.

In the Fröman and Fröman PIM, the third-order contribution is the
most important correction. Hence, it is desirable to find ways to
simplify the calculation of this contribution. For this purpose, we here
introduce a more flexible platform function $P_s(z)$ which generalizes
$\eps_0(z)$ in a way that depends on the number of singularities in
the integrand and which has the potential to simplify the evaluation
of higher-order correction terms. As a result, the choice of the so
called base function, usually denoted $Q(z)$, becomes more transparent.

\section{Higher-order phase-integral approximations and the platform function}

Let us consider the differential equation
\begin{align}
\label{eq:DE_original}
\ddzz{\psi} + R(z)\psi = 0
\end{align}
where $R(z)$ is a meromorphic function of the complex variable $z$.
Following Ref.~\onlinecite{Froman&Froman:Book:2002}, we introduce
into \eqn{eq:DE_original} a ``small'' book-keping parameter $\lambda$,
ultimately to be set equal to unity, such that we obtain the auxiliary
differential equation
\begin{align}
\label{eq:DE_auxiliary}
\ddzz{\psi} + \left(\frac{Q^2(z)}{\lambda^2} + R(z) -Q^2(z)\right)\psi = 0
\end{align}
which goes over into \eqn{eq:DE_original} when $\lambda=1$. The meromorphic
function $Q(z)$ is the unspecified base which determines the phase-integral
approximation of order one.  The auxiliary differential \eqn{eq:DE_auxiliary}
has two linearly independent solutions
\begin{subequations}
\label{eq:base}
\begin{align}
f_+ = q^{-\half}(z)\,e^{+\im w(z)} \\
f_- = q^{-\half}(z)\,e^{-\im w(z)}
\end{align}
\end{subequations}
where
\begin{align}
\label{eq:w(z)}
w(z) = \int^z q(z)\,\d{z}
\end{align}
and
\begin{align}
\label{eq:q}
& q^2(z) - q^\half(z)\ddzz{}q^{-\half}(z) 
 = R(z) - \left(1-\frac{1}{\lambda^2}\right) Q^2(z) 
\end{align}
or, alternatively,
\begin{multline}
\label{eq:q_alt}
\left(\frac{q(z)\lambda}{Q(z)}\right)^2
 - \lambda^2\frac{1}{Q^2(z)}[q(z)\lambda]^\half\ddzz{}[q(z)\lambda]^{-\half}\\
 \qquad\qquad\qquad - \lambda^2\frac{R(z)-Q^2(z)}{Q^2(z)} = 1
\end{multline}
Introducing the complex variable
\begin{align}
\label{eq:zeta}
\zeta = \int_{z_0}^z Q(z)\,\d{z}
\end{align}
where, typically, $\{z_0:Q^2(z_0)=0\}$, one finds, after some calculations,
that \eqn{eq:q_alt} can be rewritten
\begin{multline}
 \left(\frac{q(z)\lambda}{Q(z)}\right)^2
 -\lambda^2\left(\frac{q(z)\lambda}{Q(z)}\right)^\half
  \,\ddZZ{}\left(\frac{q(z)\lambda}{Q(z)}\right)^\half
\\
 - \lambda^2\frac{R(z)-Q^2(z)}{Q^2(z)}
 - \lambda^2\frac{1}{Q^{\frac{3}{2}}(z)}\ddzz{}\frac{1}{Q^\half(z)}
=1
\end{multline}
[\cf\ equation (2.2.5), with (2.2.1), in
Ref.~\onlinecite{Froman&Froman:Book:2002}].  Using \eqn{eq:zeta},
and the equality
\begin{multline}
 \frac{1}{Q^{\frac{3}{2}}(z)}\ddzz{}\frac{1}{Q^\half(z)}
 = \frac{1}{Q(z)}\ddz{}\left(\half z^s\ddz{}\frac{1}{z^sQ(z)}\right) \\
  - \left(\half z^s\ddz{}\frac{1}{z^sQ(z)}\right)^2
  + \frac{s(s-2)}{4z^2Q²(z)}
\end{multline}
this can be rewritten
\begin{multline}
\label{eq:P}
\left(\frac{q(z)\lambda}{Q(z)}\right)^2
 -\lambda^2\left(\frac{q(z)\lambda}{Q(z)}\right)^\half
  \,\ddZZ{}\left(\frac{q(z)\lambda}{Q(z)}\right)^{-\half}
\\
 + \lambda^2\left(\frac{Q^2(z)-R(z)-\frac{s(s-2)}{4z^2}}{Q^2(z)}
  - \ddZ{P_s(z)} + P_s^2(z)\right)
=1
\end{multline}
where
\begin{align}
\label{eq:platform}
P_s(z) = \half z^s\ddz{}\frac{1}{z^sQ(z)}
\end{align}
This is our new platform function. It allows us to write the right hand
member of (3.5c) in Ref.~\onlinecite{Froman&Froman:Book:1965} as
\begin{align}
\frac{R(z)-Q^2(z)}{Q^2(z)} + \ddZ{P_0(z)} -P_0^2(z)
\end{align}
In practice, the choice of $s$ in $P_s(z)$ more or less determines
the base function $Q(z)$ as we shall show next.

\section{Base function}

\begin{table}
\caption{Properties of the platform function $P_s(z)$, \eqn{eq:platform}, for
various choices of $s$.\label{tab:platform}}
\begin{tabular}{cllc} 
\toprule\toprule
$s$ &\text{Platform function} &\text{Base function} &\text{Typical $z$ range}
\\
\midrule
$0$
 &$\half \ddz{}\frac{1}{Q(z)}$
 &$Q^2(z)= R(z)$
 &$-\infty<z<\infty$
\\
$1$
 &$\half z\ddz{}\frac{1}{zQ(z)}$
 &$Q^2(z)= R(z)-\frac{1}{4z^2}$
 &$0<z<\infty$
\\
$-2l$
 &$\frac{1}{2z^{2l}}\ddz{}\frac{z^{2l}}{Q(z)}$
 &$Q^2(z)= R(z)+\frac{l(l+1)}{z^2}$
 &$0<z<\infty$
\\[.666ex]
\bottomrule\bottomrule
\end{tabular}
\end{table}

To obtain a solution of \eqn{eq:P}, we make the formal Ansatz
\begin{align}
\label{eq:formal}
\frac{q(z)\lambda}{Q(z)} = \sum_n^\infty Y_{2n}(z)\lambda^{2n}
\end{align}
and find that
\begin{subequations}
\begin{align}
Y_0 ={}& 1 \\
Y_2 = {}&-\half P_s^2 + \half\ddZ{P_s}
 - \frac{Q^2(z)-R(z)-\frac{s(s-2)}{4z^2}}{2Q^2(z)}
\end{align}
\end{subequations}

Using \eqns{eq:base} and (\ref{eq:formal}) with $\lambda=1$, and
truncating after $n=2$, we obtain the approximate local solutions of the
original differential \eqn{eq:DE_original} as
\begin{align}
\psi_\pm = \left[Q(z)\left(1+Y_2(z)\right)\right]^{-\half}
 e^{\pm\im\int^z Q(z)\left(1+Y_2(z)\right)\,\d{z}}
\end{align}
It should be noted that the integrands of these solutions contain
the total derivative
\begin{align}
\half Q(z)\ddZ{P_s(z)} = \half\ddz{P_s(z)}
\end{align}
which is integrable irrespective of the expression for $R(z)$ in the
original differential \eqn{eq:DE_original} and of the choice of
the base function $Q(z)$.  Hence, we can write
\begin{multline}
\psi_\pm = \left(\half\ddz{P_s}+\Big(1-\half P_s^2\Big)Q
 -\frac{Q^2-R-\frac{s(s-2)}{4z^2}}{2Q^2}\right)^{-\half} \\
 \times e^{\pm\im\Big[\half P_s + \int^z\Big(\big(1-\half P_s^2\big)Q
 -\frac{Q^2-R-\frac{s(s-2)}{4z^2}}{2Q^2}\Big)\,\d{z}\Big]}
\end{multline}
From this we conclude that an obvious choice for the (square of the)
base function is
\begin{align}
Q^2(z) = R(z) +\frac{s(s-2)}{4z^2}
\end{align}

For the new platform function (\ref{eq:platform}), three cases can be
discerned (see also Table~\ref{tab:platform}):
\begin{enumerate}

\item $s=0$. This corresponds to the case of an unmodified base function,
      $Q^2(z)=R(z)$. One can show that the use of the platform
      function $P_0(z)$ simplifies the calculation of phase-integral
      approximations, \eg, for the Weber functions.

\item $s=1$. This corresponds to the well-known Kramers-Langer modification
      for obtaining the base function $Q^2(z)= R(z)-1/(4z^2)$. One can
      show that the use of the platform function $P_1(z)$ simplifies the
      calculation of phase-integral approximations, \eg, for the Coulomb
      wave functions.

\item $s=-2l$. This corresponds to the omission of the centrifugal barrier
      in the base function, \ie, the choice $Q^2(z)=R(z)+l(l+1)/z^2$.
      This is (for $l\neq0$) a less known alternative; see, \eg,
      Chapter~7 in Ref.~\onlinecite{Froman&Froman:Book:1965}. One can
      show that the use of the platform function $P_{-2l}(z)$ simplifies
      the calculation of phase-integral approximations for spherically
      and cylindrically symmetric problems.

\end{enumerate}
In all the three above cases, the
third-order solution becomes
\begin{multline}
\psi_\pm = \left(\half\ddz{P_s}+\Big(1-\half P_s^2\Big)Q\right)^{-\half} \\
 \times e^{\pm\im\big[\half P_s + \int^z\big(1-\half P_s^2\big)Q\,\d{z}\big]}
\end{multline}
which exhibits the simplification attainable if the new platform function
$P_s(z)$ is used.

\section{Conclusions}

We have derived a platform function $P_s(z)$ for the third-order
correction of the phase-integral method. Since $P_s(z)$ is readily
generalized to problems involving more than one singular point in $R(z)$
and makes computational simplifications possible, this new platform
function is in many cases preferable to the conventional PIM platform
function $\eps_0(z)$.

\begin{acknowledgments}
Some of the new ideas presented in this paper were inspired by a diploma work
by Oscar Stål, who studied a model for the scattering of a radio beam
off a cylindrical ionization trail produced by ultra-high-energy
cosmic particles interacting with the Earth's lower atmosphere. One of the
authors (B.\,T.) gratefully acknowledges the financial support from the
Swedish Research Council (VR).
\end{acknowledgments}

%

\end{document}